\documentclass[%
 reprint,
showpacs,preprintnumbers,
 amsmath,amssymb,
 aps,
]{revtex4-1}
\usepackage{tipa}
\usepackage{cases}
\usepackage{stmaryrd}
\usepackage{mathrsfs}
\usepackage{amssymb}
\usepackage{amsmath}
\usepackage[english,french]{babel}

\usepackage{graphicx}
\usepackage{bm}

\begin{document}


\title{Capabilities' Substitutability and the ``S'' Curve of Export Diversity}

\author{Hongmei Lei$^{1}$}
\author{Jiang Zhang$^{1}$}\email{zhangjiang@bnu.edu.cn}

\affiliation{$^{1}$Department of Systems Science, School of Management, Beijing Normal University,100875,P.R.China}%

\date{\today}

\begin{abstract}
Product diversity, which is highly important in economic systems,
has been highlighted by recent studies on international trade. We
found an empirical pattern, designated as the ``S-shaped curve'',
that models the relationship between economic size (logarithmic GDP)
and export diversity (the number of varieties of export products) on
the detailed international trade data. As the economic size of a
country begins to increase, its export diversity initially increases
in an exponential manner, but overtime, this diversity growth slows
and eventually reaches an upper limit. The interdependence between
size and diversity takes the shape of an S-shaped curve that can be
fitted by a logistic equation. To explain this phenomenon, we
introduce a new parameter called ``substitutability'' into the list
of capabilities or factors of products in the tri-partite network
model (i.e., the country-capability-product model) of Hidalgo et al.
As we observe, when the substitutability is zero, the model returns
to Hidalgo's original model but failed to reproduce the S-shaped
curve. However, in a plot of data, the data increasingly resembles
an the S-shaped curve as the substitutability expands. Therefore,
the diversity ceiling effect can be explained by the
substitutability of different capabilities.
\end{abstract}


\maketitle


\section{Introduction}
\label{sec.introduction} Recent research on international trade has
highlighted the diversity phenomenon, which is generally ignored by
conventional economic studies. However, both the amount and the
types of goods that a country produces affect economic
growth\cite{hu.s.2011,eagle.network.2010,templet.energy.1999,krugman.increasing.1979,helpman.market.1985,petersson.export.2005,johansson.r-d.2007}.
Important new facts have been uncovered by analyzing large amounts
of high-quality data pertaining to international trade. For example,
there is a negative relationship between the diversification of
countries and the ubiquity of
products\cite{hidalgo.building.2009,hausmann.network.2011,hausmann.country.2010}.
To account for this phenomenon, Hidalgo et al. constructed a
tri-partite network model and attempted to claim that the
capabilities or non-tradable factors a country possesses are the
``building blocks'' of its economy and determine its
diversification. To their credit, the negative correlation between
diversity and ubiquity can be reproduced by their model.

However, Hidalgo et al. did not explain what ingredients determine
the non-tradable capability of a country: although they tried to
link the economic size or richness of a country with the number of
these capabilities it has on paper\cite{hausmann.country.2010}, they
did not give any empirical evidence because these capabilities are
non-measureable. In contrast, because an economy's size, as measured
by its GDP, may be the most important datum in modern economics, it
must have a correlation with a country's diversification
degree\cite{zhang.allometric.2010}. It is obvious that countries
with large GDP always produce and export more diversified products
and that countries with small GDP usually have more homogenous
products and
markets\cite{sozen.prediction.2007,mozumder.causality.2007,camba-mendez.automatic.2001}.
This observation can be described quantitatively by an S-shaped
curve that models a country's logarithmic GDP and export diversity
\cite{hu.s.2011,haanstra.use.1985,zwietering.modeling.1990,northam.urban.1979}.
This interdependence between size and diversity is ubiquitous in
global trade and economic systems; furthermore, it is common in
ecological
systems\cite{hubbell.unified.2008,gaston.biodiversity.2004,mulder.physical.2001,loreau.biodiversity.2001}.
The classical ``area-species'' relationship in ecology which is
another example of the interdependence between size and diversity
resembles an S-shaped
curve\cite{hu.s.2011,he.species-area.1996,he.spatial.1996,wei.comparative.2010}.

To this point, the theoretical understanding of the S-shaped curve
as a model of the relationship between economic size and export
diversity is still deficient. Therefore, this paper tries to build a
model to reproduce this size-diversity curve. Initially, we simply
link the probability that a country may possess certain capabilities
with its economic size. In this way, we can investigate how economic
size determines a country's diversification. However, this
interdependence between size and diversity in Hidalgo et al.'s model
is exponential, as they have noted; thus, once a country's economic
size exceeds a certain threshold, the country will receive
increasing returns. As a result, the type of products the country's
businesses are able to export increases without any limitation;
otherwise, the country's economy cannot overcome the so-called
``quiescence
trap''\cite{hausmann.country.2010,hausmann.network.2011}. However,
the empirical data reveal that there is an upper limit of the
diversity curve\cite{hu.s.2011} that cannot be reproduced by the
original tri-partite model. Therefore, we have introduced an
important parameter into our model, namely, the substitutability $s$
between different capabilities; this parameter's purpose is to relax
the overly strict condition on the number of capabilities that a
product requires. Interestingly, in
paper\cite{hausmann.country.2010}, Hidalgo et al. mentioned the idea
of substitutability between factors but they did not develop it. In
this paper, we report that our model that includes the
substitutability $s$ can reproduce the S-shaped curve of economic
diversity.

This paper is organized as follows: in Section \ref{sec.method}, we
briefly introduce the S-shaped relationship between the export
diversity and economic size of countries into our model to simulate
the empirical relationship between these factors and to achieve both
accurate and approximate analytic solutions. In
Section\ref{sec.results}, we show the simulation and analytic
results (both exactly and approximately), which can resemble the
empirical ``S'' curve. Furthermore, we discuss how the key
parameters in our model affect the shapes of fitting curves.

\section{Method}
\label{sec.method}

\subsection{The S-shaped relationship}
The S-shaped relationship between logarithmic GDP and export
diversity can be derived from the empirical data we have collected.
The world GDP statistics are from the World Bank's web-site
(www.worldbank.org) and the export diversity data are from the
NBER-UN world trade database (www.nber.org/data). In the former
data-set, information including the GDPs, populations and other
economic data from 240 countries was recorded during 1971-2006; in
the latter data set, the detailed bilateral trade flows of
approximately 150 countries and 800 types of products (according to
the SITC4 classification standard) during 1962-2000 are included. In
this paper, we only show the ``S'' curve in 1995. A more detailed
discussion of empirical ``S'' curves for other years can be gleaned
from previous work \cite{hu.s.2011}.

\begin{center}
\begin{figure*}[!ht]
\includegraphics[scale=0.7]{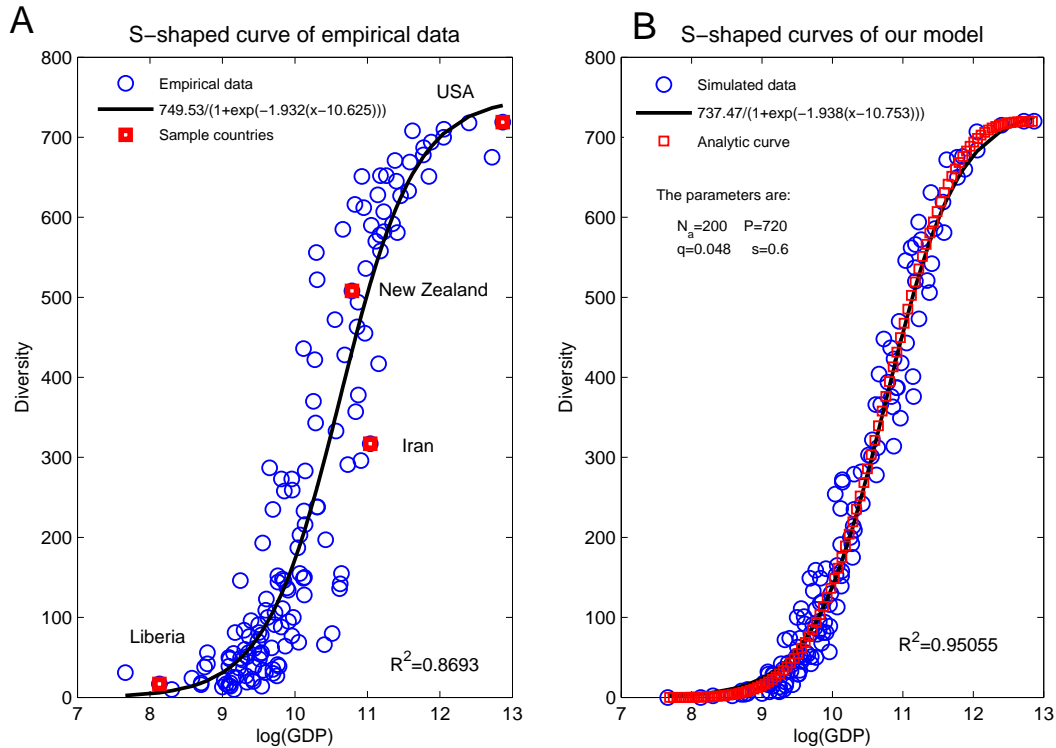} \caption{The S-shaped relationship between
log(GDP) ($X_{i}$) and the export diversity $D_{i}$ of countries.
(A) The original data and the logistic fitting for all countries in
1995. (B) The blue scatter points are the simulation results from
our tri-partite model. Whereas the red curve is the analytic result
of Equation \ref{complexprob2}(see section \ref{sec.model} of main
text), the black curve is the logistic fitting.} \label{fig.scurve}
\end{figure*}
\end{center}

The empirical data shows a strong dependence between logarithmic GDP
and the types of exports in FIG \ref{fig.scurve}A ($R^2=0.87$). The
empirical data can be fitted by a logistic function; note that such
functions are widely used in many
disciplines\cite{zwietering.modeling.1990,wei.comparative.2010,scheiner.six.2003}:

\begin{equation}
\label{logisticfunction} D_i=\frac{A}{1+e^{-k(X_i-x_m)}}
\end{equation}

where, $D_i$ stands for the number of categories of goods (each
category is represented by a distinct 4-digits code) that country
$i$ export and $X_i$ represents the logarithmic GDP of country $i$.
Furthermore, $A,k$ and $x_m$ are parameters of the logistic
function. The estimated values are shown in the legend of FIG
\ref{fig.scurve}A.

From FIG.\ref{fig.scurve}, we observe that all countries can be
divided into three groups: small countries that are in the lower
part of the ``S'' curve (e.g., Liberia), intermediate countries
(e.g., Iran and New Zealand) that are in the ``accelerate'' part of
the curve and large countries (e.g., the USA) that are at the top of
the curve. The small countries in the first group can export very
few varieties of goods if their sizes do not exceed a certain
threshold (the ``quiescence trap''\cite{hausmann.country.2010}). The
second segment of the ``S'' curve represents an exponential
increase, in which, countries have very different types of export
products. However, the accelerating-growth effect stops at the
uppermost part of the ``S'' curve due to the ceiling effect of
diversity, as large countries' diversification levels are not as
high as an exponential curve would indicate. This pattern in the
relationship between economic size and export diversity is very
stable in all of the years of our data set (see \cite{hu.s.2011}).

\subsection{Model}
\label{sec.model}

It is important to consider why this interdependence between size
and diversity in international trade exits. In paper
\cite{hidalgo.building.2009}, Hidalgo et al. proposed a tri-partite
network model to account for various facts regarding export
diversity and product ubiquity. In their model, the first and third
layered nodes are the countries and their products, respectively,
whereas the nodes in the hidden layer between countries and products
are introduced to represent non-tradable elements, factors or
capabilities, e.g., management skills, raw materials, regulation,
property rights, etc. Thus, countries need to have these elements
locally available to produce goods.

Following Hidalgo's model, we hope to account for the S-shaped
relationship by constructing a modified model which also assumes
that each product requires some non-trade capabilities, and each
country can export a product if and only if this country possesses
the required capabilities.

\begin{figure}
\centering
\includegraphics[scale=0.6]{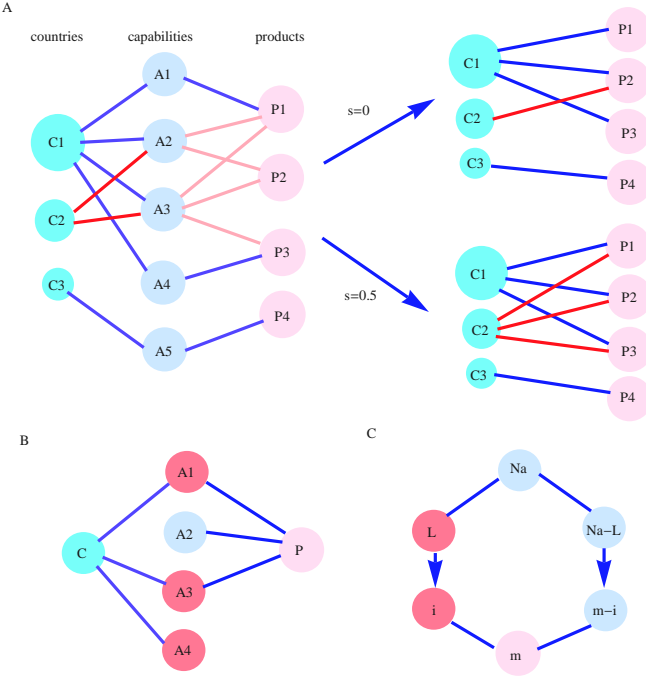}
\caption{A concept graph of three-layer network and the analytic
solution. (A) This three-layer network is designed to depict the
mapping between countries and products. The nodes in the first layer
represent countries and the nodes' sizes represent logarithmic their
GDPs. The nodes in the second and third layers represent the
capabilities and products, respectively. A country can produce more
products when the substitutability $s$ increases. (B) Provided that
country $c$ has L distinct capabilities (such capabilities are
colored in red), the probability that this country produces product
$p$ is determined by the mapping between these capabilities and
product $p$. (C) The $m$ capabilities required by product $p$ can be
divided into two groups based on whether they match the capabilities
that are owned by country $c$.} \label{fig.sketchmap}
\end{figure}

However, we initially link the logarithmic GDP with the degrees of
the nodes that represents countries because our purpose is to
explain the relationship between economic size and economic
diversity. That is, the number of links to country $c$  is
proportional to its logarithmic GDP $X_{c}$, but these links'
end-nodes are randomly selected among the capability nodes that
represents a country $i$ that possesses the given capability (see
FIG.\ref{fig.sketchmap}A). Furthermore, the links between
capabilities and products are randomly assigned except that they are
constrained by the given connection density $q$. The tri-partite
network is constructed in this way.

In Hidalgo et al.'s model, country $c$ can export product $p$ if and
only if the paths from $c$ to $p$ include all of the hidden nodes
that connect $c$ to $p$. Thus, all of the capabilities that are
devoted to producing $p$ are possessed by country $i$. However, this
mechanism cannot reproduce the ``S'' curve that models the
relationship between size and diversity, and as a result, we must
replace this mechanism with a new rule we have designed.

We introduce one important parameter called the ``average
substitutability rate'' (or ``substitutability'') $s$ to represent
the proportions of the total capabilities that are required to
produce product $p$£»; these capabilities can be replaced by other
available capabilities. Country $c$ can export product $p$ if and
only if the paths from $c$ to $p$ cover $(1-s)*100\%$ of the
capabilities required by $p$ (which would imply that the hidden
nodes are connected to $p$). Hence, among the requisite
capabilities, only $1-s$ fractions are necessary and other $s$
fractions are substitutable in average. When the substitutability
$s$ is 0, all of the capabilities are necessary, and as a result, we
recover Hidalgo et al.'s model. However, when $s$ increases, more
countries can export diverse products to the same extent as the
largest countries. As a result, an S-shaped curve between
logarithmic GDP and diversity is obtained.

For example (see FIG. \ref{fig.sketchmap}A), suppose that country C2
can only export product P2 when $s = 0$ and that C2 cannot export P1
because all of the capabilities connected to P1 (namely, A1, A2, and
A3) would have to be covered by the paths from C2 to P2, yet A1 is
not covered. When $s$ increases to 0.5, only 50\% of the
capabilities must be covered by the paths. Therefore, C2 can export
P1 because more than half of the capabilities required by P2 have
been covered.

In general, we consider N countries whose logarithmic GDPs are
($X_{1},X_{2},\cdot\cdot\cdot,X_{N}$). The adjacency matrix between
countries and capabilities is $C_{ik}$, and its elements are
randomly assigned:

\begin{subnumcases}
{C_{ik}=}
1, &with probability $r_i$\\
0, &otherwise,
\end{subnumcases}\label{countryability}

where the subscript $k$ is iterated from 1 to $N_{a}$ (which is the
total number of capabilities we consider). To link the number of
capabilities that a country has with this country's GDP, we assume
that the probability $r_{i}$ is proportional to the value of $X_{i}$
of country $c_{i}$ symbolically,

\begin{equation}
{r_i \propto \log(X_{i})}
\end{equation}\label{countryprob}

Actually, any linear relationship between $r_{i}$ and $X_{i}$ can
produce an S-shaped curve. In our model, we let
$r_{i}=(X_{i}-X_{m})/(X_{M}-X_{m})$ to reduce the number of
parameters as much as possible, where $X_{M}$ and $X_{m}$ are the
largest and smallest value of log(GDP) in the list of countries,
respectively. Additionally, we assign connections from product
$p_{j}$ to the required capabilities with probability $q$. The
matrix $P_{kj}$ represents the connections between these two layers:

\begin{subnumcases}
{P_{kj}=}
1, &with probability $q$\\
0, &otherwise.
\end{subnumcases}\label{productability}

In the above equation, $j$ is iterated from 1 to $P$ (the total
number of possible products). Suppose the adjacency matrix between
countries and products is $M_{ij}$. We assume that country $c_{i}$
can export product $p_{j}$( i.e., that $M_{ij}=1$), if and only if
the proportion of capabilities that are owned by the producing
countries is at least a certain percent ($s$) of the total number of
capabilities that are required by products:

\begin{subnumcases}
{M_{ij}=}
1, &if ${\sum_{k}}C_{ik}P_{kj} \geq (1-s){\sum_{k}}P_{kj}$\\
0, &otherwise.
\end{subnumcases} \label{newcountryproduct}

Finally, the export diversity $D_{i}$ is defined as the total number
of types of products that country $c_{i}$ exports, namely,

\begin{equation}
\label{totalnumber} {D_{i}=}{\sum_{j}M_{ij}}.
\end{equation}

In the simulations, all $X_{i}$s are defined by the real-world
log(GDP) data that we have collected, the number of countries is
$N$, and $P$ is defined as the total number of products in our
data-set. The number of capabilities $N_{a}$, the link density
between the capabilities and products $q$, and the substitutability
rate $s$ are the parameters. In each simulation, we can generate a
tri-partite network according to the rules we introduced, and as a
result, the relationship between $X_{i}$ and $D_{i}$ can be derived.

\subsection{Analytic solution}

Before giving the simulation results, we will first derive the
analytic relation between $D_{i}$ and $X_{i}$ to explain the
mathematical essence of this model can be grasped.

In our model, all of the connections among the countries,
capabilities and products are independent \emph{per se}. Therefore,
by analyzing the probability that a typical country $c_{i}$ exports
a specific product $p_{j}$, we can derive its export diversity:

\begin{equation}\label{singlediversity}
D_i=P \pi_i.
\end{equation}

where, $\pi_i$ is the probability that country $c_{i}$ can produce
any specific product.

Suppose the capabilities that country $c_{i}$ possesses and that
product $p_{j}$ requires are $k_{c_i}$ and $k_{p_j}$, respectively.
Because the density of capability that a country has is proportional
to $X_{i}$, the expectation value of $k_{c_i}$ is also proportional
to $X_{i}$. To simplify our discussion, we treat $k_{c_i}$ as a
predetermined value and use $k_{c_i}$ to represent its expectation
value.

The probability $\pi_i$ can be decomposed into various other
probabilities as follows:

\begin{equation}\label{normprob}
\pi_{i}={\sum_{m=0}^{N_{a}}}\Pr\{k_{p_{j}}=m\}\cdot
\Pr\{c_i\rightarrow p_j|k_{p_j}=m\},
\end{equation}

where $\Pr\{c_i\rightarrow p_j|k_{p_j}=m\}$ is the probability that
country $c_{i}$ exports product $p_{j}$, which depends on the degree
($k_{p_j}$) of $p_{j}$ being $m$.

First, because a product has a probability q of requiring one
capability, the number of capabilities $k_{p_j}$ required by $p_{j}$
satisfies a binomial distribution. Hence, we know the probability
that node $p_{j}$ requires $m$ distinct capabilities is:

\begin{equation}
\label{probprodct}
\Pr\{(k_{p_{j}}=m)\}=\binom{N_{a}}{m}q^m(1-q)^{N_{a}-m}.
\end{equation}

Second, we derive $\Pr\{c_i\rightarrow p_j|k_{p_j}=m\}$. If the
number of connections of nodes $c_{i}$ and $p_{j}$ are given, then
the situation can be as depicted by FIG.\ref{fig.sketchmap}B. The
probability $\Pr\{c_i\rightarrow p_j|k_{p_j}=m\}$ is the number of
connection configurations satisfying that the number of elements in
the set of capabilities that are connected with both $c_{i}$ and
$p_{j}$ is larger than $(1-s)m$ over all of the possible connection
configurations. This number is computed by means of the following
steps:

i) There are $N_{a}(N_{a}-1)( N_{a}-2)\cdots(N_{a}-m+1)$(i.e.,
permutation $P_{N_{a}}^m$) ways that the product $p_{j}$ is
connected to $m$ capabilities.

ii) All of the $m$ capabilities that are required by product $p_{j}$
can be divided into two groups based on whether they are owned by
country $c_{i}$. Without loss of generality, suppose there are $n$
capabilities in the first group (that are possessed by $c_{j}$,
i.e., $A3$ and $A4$ in FIG.\ref{fig.sketchmap}B), and $m-n$
capabilities in the second group (i.e., $A2$ in
FIG.\ref{fig.sketchmap}B). (See FIG.\ref{fig.sketchmap}C.)

iii) There are
$P_{k_{c_i}}^n=k_{c_i}(k_{c_i}-1)(k_{c_i}-2)\cdots(k_{c_i}-n+1)$
ways to match the capabilities in the first group.

iv) Similar to iii), the number of ways that the capabilities in the
second group can be matched with the $k_{c_i}-n$ capabilities that
are not owned by country $c_{i}$ is $P_{N_{a}-k_{c_j}}^{m-n}$.

v) There are $\binom{m}{n}$ ways to select $n$ elements from $m$
capabilities.

Indeed, because there must be at least $[(1-s)m]$ capabilities in
the first group,so we can obtain:

\begin{equation}
\label{complexprob1} \Pr\{c_i\rightarrow
p_j|k_{p_j}=m\}={\sum_{n=n_1}^{m}}\frac{\binom{m}{n} P_{k_{c_i}}^n
P_{N_{a}-k_{c_i}}^{m-n}}{P_{N_{a}}^m}.
\end{equation}

In the above equation, the summation index $n$ begins at
$n_1=\max([(1-s)m],k_{c_i}+m-N_{a})$ because the number of
capabilities ($k_{c_i}-n$) that are not owned by $c_{i}$ cannot
exceed $N_a-m$ and $n$ must be larger than $k_{c_i}+m-N_a$. By
inserting Equations \ref{complexprob1} and \ref{probprodct} into
Equations \ref{normprob} and \ref{singlediversity}, we can derive
the following:

\begin{equation}
\label{complexprob2}
D_i=P{\sum_{m=0}^{N_{a}}}\binom{N_{a}}{m}q^m(1-q)^{N_{a}-m}\sum_{n=n_1}^{m}\frac{\binom{m}{n}
P_{k_{c_i}}^n P_{N_{a}-k_{c_i}}^{m-n}}{P_{N_{a}}^m}.
\end{equation}

Notice that $k_{c_i}\propto X_i$, which implies that $D_{i}$ is
actually a function of $X_{i}$.

Although Equation \ref{complexprob2} accurately models the relation
between $X_{i}$ and $D_{i}$, it is complex; however, we can simplify
it to an approximate but compact form. If we allow duplicate links
to exist in the network, then the permutations in Equation
\ref{complexprob2} can be replaced by exponentials, and thus, each
permutation $P_x^y$ can be replaced with $x^y$. Furthermore, we can
use $[(1-s)m]$ to approximate $\max([(1-s)m],k_{c_i}+m-N_a)$; then,
we have

\begin{equation}
\label{easyprob1}
\begin{aligned} D_i\approx
P\sum_{m=0}^{N_{a}}\binom{N_{a}}{m}
&q^m(1-q)^{N_{a}-m}\cdot\\&\sum_{n=[(1-s)m]}^m\binom{m}{n}(\frac{k_{c_i}}{N_a})^n(1-\frac{k_{c_i}}{N_a})^{m-n}.
\end{aligned}
\end{equation}

When $s=0$, Equation \ref{easyprob1} becomes
$(\frac{k_{c_i}q}{N_a}+1-q)^{N_a}$ according to the binomial
theorem. Because Equation 12 is the same equation as the relation
between capability and diversity derived that was in paper
\cite{hausmann.country.2010}, Equation \ref{easyprob1} is actually a
general definition of $D_{i}$ in terms of $X_{i}$ in which
substitutability between capabilities is allowed.

\section{Results}
\label{sec.results}

\subsection{The S-shaped curve}

In the previous sections we introduced our model. Here, we will give
our simulation and numeric results.

In FIG.\ref{fig.scurve}B, the blue circles represent the simulation
results and the red squares represent both the numeric results of
Equation \ref{easyprob1} and the logistic fitting. When we set the
number of capabilities ($N_{a}$) at 200
\cite{hidalgo.building.2009}, the number of products ($P$) at 720
(which is also the maximum product diversity of the countries in our
empirical data), the link density of capabilities and products ($q$)
at 0.048, and the substitutability at 0.6, we obtain an ``S'' curve
that resembles the empirical curve of best fit for the data recorded
in 1995. Furthermore, we use the logistic Equation
\ref{logisticfunction} to fit both the empirical and theoretical
curves and to compare their fitting parameters. We found that the
parameters are similar: whereas $A=749.53,k=1.932,X_{M}=10.625$ for
the empirical curve, $A=737.47,k=1.938,X_{M}=10.753$ for the
theoretical curve. Therefore, we conclude that our model can
simulate the empirical S-shaped relationship very well.

\subsection{Parameter Space}

\begin{center}
\begin{figure*}[!ht]
\centering
\includegraphics[scale=0.9]{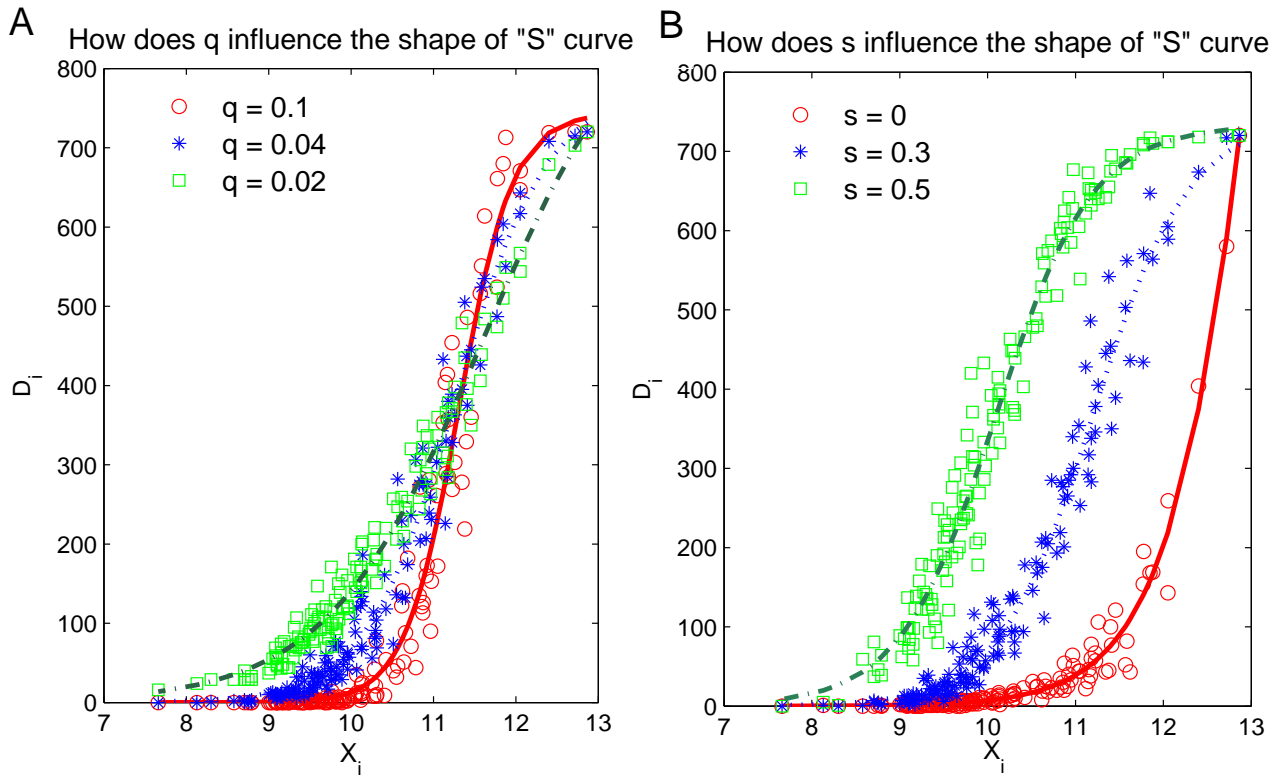} \caption{The S-shaped relationship, which depends on
our model's parameters. (A) These S-shaped curves change if the
parameter $q$ changes and the other parameters are kept fixed; when
$q$ increases, the S-shaped curve becomes steeper and the trade gap
among countries increases. (B) These S-shaped curves change if the
parameter $s$ changes and the other parameters are kept fixed. The
ceiling of the S-shaped curve emerges when $s$ increases
sufficiently.} \label{fig.qandr}
\end{figure*}
\end{center}

Although there are several parameters, the most important ones are
$q$ and $s$. In fact, we can fix the other parameters (specifically,
we select $P=720$ and $N_{a}=200$) and study how $q$ and $s$ affect
the shape of the ``S'' curve. From the notions introduced above, the
parameter $q$ determines the capabilities that are required by
products. Thus, we can understand $q$ as the average complexity of
all products. As $q$ increases, countries find it more difficult to
make products, and as a result, the S-shaped curve is steeper and
the diversity gap between rich countries and poor countries becomes
large (see FIG.\ref{fig.qandr}A and
\cite{hidalgo.product.2007,ausloos.clusters.2007}).

The parameter $s$ represents the average substitutability degree of
the products: one country must possess a proportion of $1-s$ of the
capabilities required by a product if this country wants to export
that product. From FIG.\ref{fig.qandr}B, no ceiling for the S-shaped
curve can be observed when $s$ is small because countries need to
have locally almost all of the capabilities required by products in
this case. Thus, when $s$ is zero, the simulation result is the same
as the result of Hidalgo's model. In contrast, a ceiling for the
export diversity emerges as $s$ increases because more capabilities
that are required by products can be replaced by other available
capabilities that are owned by producing countries. Hence, the
resources (for instance, labor, skills, fund and so on) required to
produce the goods in question are more diverse.

\begin{figure}
\centering
\includegraphics[scale=0.6]{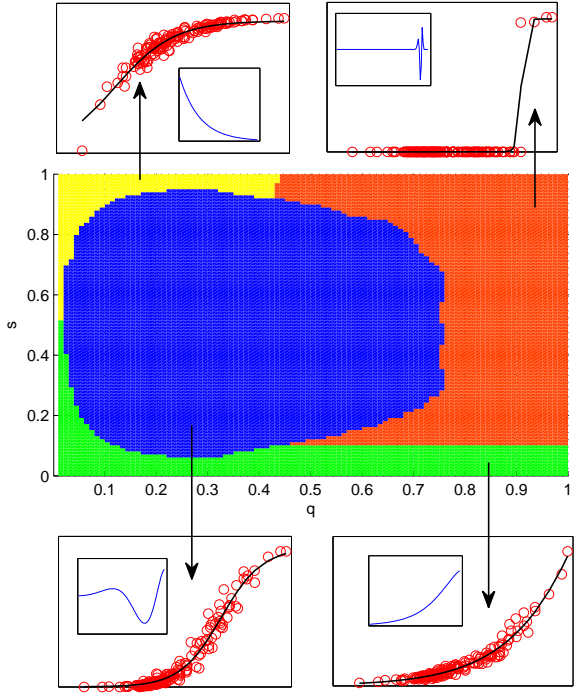}
\caption{Different curves and their third-order derivatives in
parameter space. The different regions of parameters $q$ and $s$
with different colors represent different shapes of the numeric
curves in our model. Additionally, the small figures plot the curves
between $X_{i}$ and $D_{i}$ and the insets show their third-order
derivatives.} \label{fig.figurepart}
\end{figure}

More numeric experiments are implemented to investigate how the
shape of the ``S'' curve changes with changes in the combinations of
$q$ and $s$. The results show that the parameter space (i.e., the
combinations of $q$ and $s$) can be decomposed into several regions,
as shown in FIG.\ref{fig.figurepart}. The blue region in
FIG.\ref{fig.figurepart} represents the combinations of $q$ and $s$
that can generate a curve that models the relationship between size
and diversity and that exhibits an obvious ``S'' shape. However, the
curves in every parameter region except for the blue one have the
shape of twisted ``S'' curves only partially. We can distinguish
these regions by considering the third- order derivatives
($d^{(3)}D_i/dX_i^3$): if the third-order-derivative curve
$d^{(3)}D_i/dX_i^3$ can be separated by the x-axis into three
segments, then the original curve is clearly S-shaped. However, if
the curve of $d^{(3)}D_i/dX_i^3$ has only one or two segments that
are divided by the x-axis, then the original curves are not
S-shaped.

\begin{figure}
\centering
\includegraphics[scale=0.8]{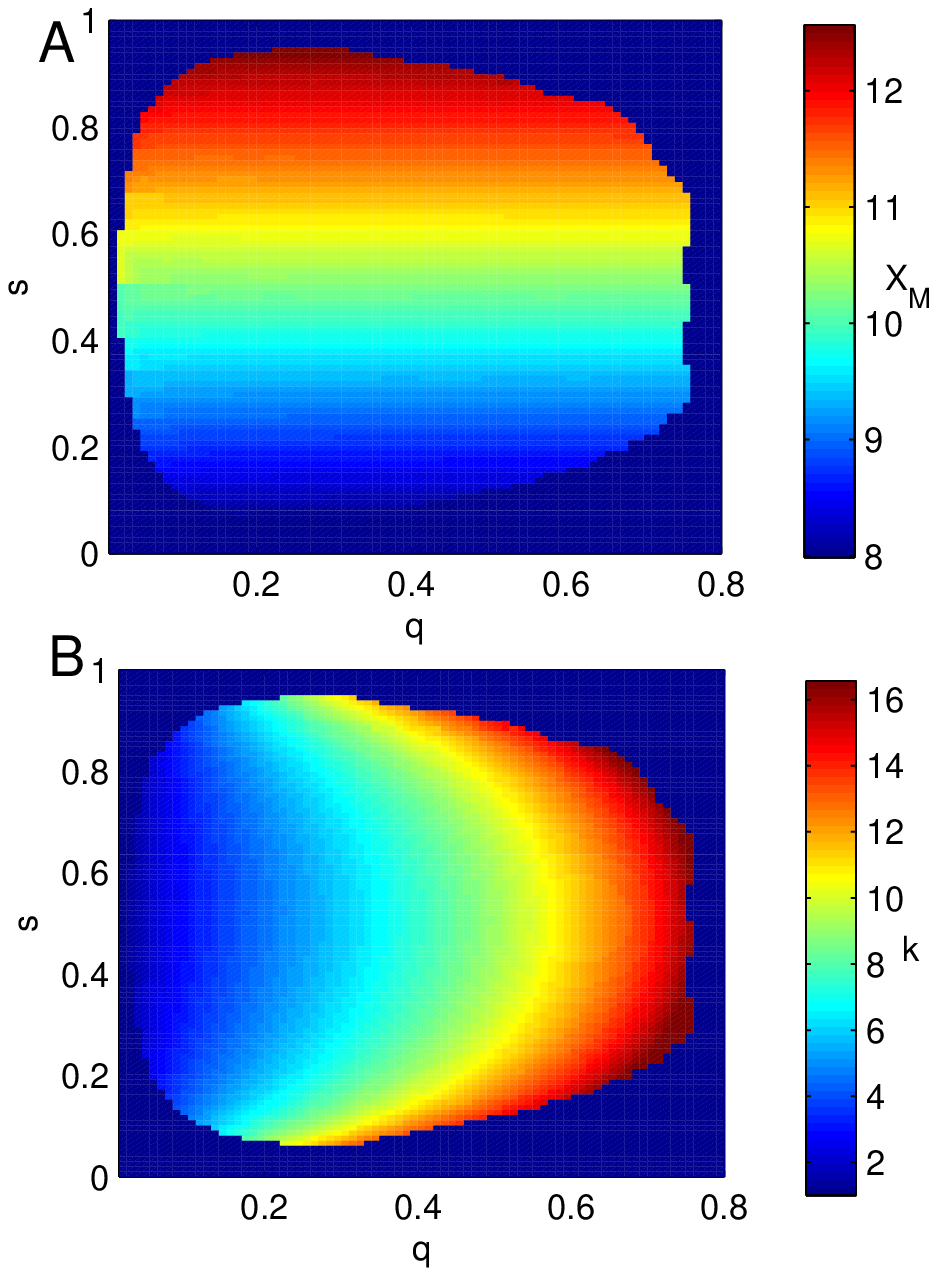}
\caption{The parameters $X_{M}$ (A) and $k$ (B) of the logistic
function depend on the parameter $q$ and $s$ in our model. Only the
parameter regions that can generate ``S'' curves are shown; the blue
areas are blank.} \label{fig.KandX_M}
\end{figure}

To quantitatively characterize the curves that model the
relationship between $X_{i}$ and $D_{i}$, we use the logistic
function (Equation \ref{logisticfunction}) to fit the curves and
show how the parameters (i.e., $k$ and $x_{m}$) change when the
combinations of $q$ and $s$ are varied (FIG.\ref{fig.KandX_M}).
However, we only show the regions of $q$ and $s$ that will generate
a stable ``S'' shape because the logistic fitting would otherwise
give unreasonable fitting parameters. From FIG.\ref{fig.KandX_M}, we
can observe that whereas the slope ($k$) of $X_i$-$D_i$ is
influenced mainly by $q$ and not $s$, the center position of the
curve is determined mainly by $s$.

\section{Discussion}
In general, this paper discusses how the revision of Hidalgo et
al.'s tri-partite network model can generate the observed S-shaped
curve of the global export diversity, which depends on the economic
sizes of countries. In this model, we found that the
substitutability $s$ is an important parameter that can account for
the ceiling effect in the S-shaped curve. When $s$ decreases, the
size - diversity curve increasingly resembles a logistic curve and
becomes dissimilar to the exponential function predicted by paper
\cite{hausmann.country.2010}. Therefore, we claimed the
substitutability between different capabilities cannot be ignored
because the empirical size - diversity curve has an S-shape.

However, this work is only the first step toward a fuller
understanding of the export diversity in international trade. The
S-shaped curve that models the relationship between diversity and
economic size can only show the aggregate information regarding one
country's export diversity. Additional studies that investigate the
distribution of different products in a given country are worth
conducting in future.

\begin{acknowledgements}
We acknowledge the support off the National Natural Science
Foundation of China under Grant No.61004107, and we are grateful for
our discussions with Professors Y.G. Wang, W.X. Wang and Prof. Q.H.
Chen. In addition, we acknowledge the excellent advice of Deli He.
\end{acknowledgements}

\bibliography{references}

\begin{thebibliography}{27}%
\makeatletter
\providecommand \@ifxundefined [1]{%
 \@ifx{#1\undefined}
}%
\providecommand \@ifnum [1]{%
 \ifnum #1\expandafter \@firstoftwo
 \else \expandafter \@secondoftwo
 \fi
}%
\providecommand \@ifx [1]{%
 \ifx #1\expandafter \@firstoftwo
 \else \expandafter \@secondoftwo
 \fi
}%
\providecommand \natexlab [1]{#1}%
\providecommand \enquote  [1]{``#1''}%
\providecommand \bibnamefont  [1]{#1}%
\providecommand \bibfnamefont [1]{#1}%
\providecommand \citenamefont [1]{#1}%
\providecommand \href@noop [0]{\@secondoftwo}%
\providecommand \href [0]{\begingroup \@sanitize@url \@href}%
\providecommand \@href[1]{\@@startlink{#1}\@@href}%
\providecommand \@@href[1]{\endgroup#1\@@endlink}%
\providecommand \@sanitize@url [0]{\catcode `\\12\catcode `\$12\catcode
  `\&12\catcode `\#12\catcode `\^12\catcode `\_12\catcode `\%12\relax}%
\providecommand \@@startlink[1]{}%
\providecommand \@@endlink[0]{}%
\providecommand \url  [0]{\begingroup\@sanitize@url \@url }%
\providecommand \@url [1]{\endgroup\@href {#1}{\urlprefix }}%
\providecommand \urlprefix  [0]{URL }%
\providecommand \Eprint [0]{\href }%
\@ifxundefined \urlstyle {%
  \providecommand \doi  [0]{\begingroup \@sanitize@url \@doi}%
  \providecommand \@doi [1]{\endgroup \@@startlink {\doibase
  #1}doi:\discretionary {}{}{}#1\@@endlink }%
}{%
  \providecommand \doi  [0]{doi:\discretionary{}{}{}\begingroup
  \urlstyle{rm}\Url }%
}%
\providecommand \doibase [0]{http://dx.doi.org/}%
\providecommand \Doi [0]{\begingroup \@sanitize@url \@Doi }%
\providecommand \@Doi  [1]{\endgroup\@@startlink{\doibase#1}\@@Doi}%
\providecommand \@@Doi [1]{#1\@@endlink}%
\providecommand \selectlanguage [0]{\@gobble}%
\providecommand \bibinfo  [0]{\@secondoftwo}%
\providecommand \bibfield  [0]{\@secondoftwo}%
\providecommand \translation [1]{[#1]}%
\providecommand \BibitemOpen [0]{}%
\providecommand \bibitemStop [0]{}%
\providecommand \bibitemNoStop [0]{.\EOS\space}%
\providecommand \EOS [0]{\spacefactor3000\relax}%
\providecommand \BibitemShut  [1]{\csname bibitem#1\endcsname}%

\bibitem [{\citenamefont {Hu}\ \emph {et~al.}(2011)\citenamefont {Hu},
  \citenamefont {Tian}, \citenamefont {Wang},\ and\ \citenamefont
  {Zhang}}]{hu.s.2011}%
  \BibitemOpen
  \bibfield  {author} {\bibinfo {author} {\bibfnamefont {L.}~\bibnamefont
  {Hu}}, \bibinfo {author} {\bibfnamefont {K.}~\bibnamefont {Tian}}, \bibinfo
  {author} {\bibfnamefont {X.}~\bibnamefont {Wang}}, \ and\ \bibinfo {author}
  {\bibfnamefont {J.}~\bibnamefont {Zhang}},\ } {\bibfield  {journal} {\bibinfo  {journal}
  {Physica A: Statistical Mechanics and its Applications},\ }\textbf {\bibinfo {volume} {391}},\ \bibinfo {pages} {731-739} (\bibinfo {year} {2012})}%
\bibitem [{\citenamefont {Eagle}\ \emph {et~al.}(2010)\citenamefont {Eagle},
  \citenamefont {Macy},\ and\ \citenamefont {Claxton}}]{eagle.network.2010}%
  \BibitemOpen
  \bibfield  {author} {\bibinfo {author} {\bibfnamefont {N.}~\bibnamefont
  {Eagle}}, \bibinfo {author} {\bibfnamefont {M.}~\bibnamefont {Macy}}, \ and\
  \bibinfo {author} {\bibfnamefont {R.}~\bibnamefont {Claxton}},\ }{\bibfield  {journal} {\bibinfo  {journal}
  {Science},\ }\textbf {\bibinfo {volume} {328}},\ \bibinfo {pages} {1029}
  (\bibinfo {year} {2010})},\ ISSN \bibinfo {issn} {0036-8075,
  1095-9203}%
  
\bibitem [{\citenamefont {Templet}(1999)}]{templet.energy.1999}%
  \BibitemOpen
  \bibfield  {author} {\bibinfo {author} {\bibfnamefont {P.~H.}\ \bibnamefont
  {Templet}},\ }{\bibfield  {journal}
  {\bibinfo  {journal} {Ecological Economics},\ }\textbf {\bibinfo {volume}
  {30}},\ \bibinfo {pages} {223} (\bibinfo {year} {1999})},\ ISSN \bibinfo
  {issn} {0921-8009}%
  

\bibitem [{\citenamefont {Krugman}(1979)}]{krugman.increasing.1979}%
  \BibitemOpen
  \bibfield  {author} {\bibinfo {author} {\bibfnamefont {P.~R.}\ \bibnamefont
  {Krugman}},\ }{\bibfield  {journal}
  {\bibinfo  {journal} {Journal of International Economics},\ }\textbf
  {\bibinfo {volume} {9}},\ \bibinfo {pages} {469} (\bibinfo {year} {1979})},\
  ISSN \bibinfo {issn} {0022-1996}%
\bibitem [{\citenamefont {Helpman}\ and\ \citenamefont
  {Krugman}(1985)}]{helpman.market.1985}%
  \BibitemOpen
  \bibfield  {author} {\bibinfo {author} {\bibfnamefont {E.}~\bibnamefont
  {Helpman}}\ and\ \bibinfo {author} {\bibfnamefont {P.~R.}\ \bibnamefont
  {Krugman}},\ }{{\selectlanguage {english}\emph {\bibinfo {title}
  {Market Structure and Foreign Trade: Increasing Returns, Imperfect
  Competition, and the International Economy}}}}\ (\bibinfo  {publisher} {{MIT}
  Press},\ \bibinfo {year} {1985})\ ISBN \bibinfo {isbn}
  {9780262580878}%
\bibitem [{\citenamefont {Petersson}(2005)}]{petersson.export.2005}%
  \BibitemOpen
  \bibfield  {author} {\bibinfo {author} {\bibfnamefont {L.}~\bibnamefont
  {Petersson}},\ } {\bibfield  {journal}
  {\bibinfo  {journal} {The South African Journal of Economics},\ }\textbf
  {\bibinfo {volume} {73}},\ \bibinfo {pages} {785} (\bibinfo {year} {2005})},\
  ISSN \bibinfo {issn} {0038-2280, 1813-6982}%
\bibitem [{\citenamefont {Johansson}\ and\ \citenamefont
  {Karlsson}(2007)}]{johansson.r-d.2007}%
  \BibitemOpen
  \bibfield  {author} {\bibinfo {author} {\bibfnamefont {S.}~\bibnamefont
  {Johansson}}\ and\ \bibinfo {author} {\bibfnamefont {C.}~\bibnamefont
  {Karlsson}},\ }{\bibfield
  {journal} {\bibinfo  {journal} {The Annals of Regional Science},\ }\textbf
  {\bibinfo {volume} {41}},\ \bibinfo {pages} {501} (\bibinfo {year}
  {2007})}%
\bibitem [{\citenamefont {Hidalgo}\ and\ \citenamefont
  {Hausmann}(2009)}]{hidalgo.building.2009}%
  \BibitemOpen
  \bibfield  {author} {\bibinfo {author} {\bibfnamefont {C.~A.}\ \bibnamefont
  {Hidalgo}}\ and\ \bibinfo {author} {\bibfnamefont {R.}~\bibnamefont
  {Hausmann}},\ } {\bibfield  {journal} {\bibinfo
   {journal} {Proceedings of the National Academy of Sciences of the United
  States of America},\ }\textbf {\bibinfo {volume} {106}},\ \bibinfo {pages}
  {10570} (\bibinfo {year} {2009})},\ ISSN \bibinfo {issn} {0027-8424},\
  \bibinfo {note} {{PMID:} 19549871 {PMCID:} {PMC2705545}}%
\bibitem [{\citenamefont {Hausmann}\ and\ \citenamefont
  {Hidalgo}(2011)}]{hausmann.network.2011}%
  \BibitemOpen
  \bibfield  {author} {\bibinfo {author} {\bibfnamefont {R.}~\bibnamefont
  {Hausmann}}\ and\ \bibinfo {author} {\bibfnamefont {C.~A.}\ \bibnamefont
  {Hidalgo}},\ }{\bibfield  {journal} {\bibinfo  {journal}
  {Journal of economic growth. - Dordrecht [u.a.] : Springer Science Business
  Media Inc., {ISSN} 1381-4338, {ZDB-ID} 13143827. - Vol. 16.2011, 4, p.
  309-342}} (\bibinfo {year} {2011})}%
\bibitem [{\citenamefont {Hausmann}\ \emph {et~al.}(2010)\citenamefont
  {Hausmann}, \citenamefont {Hidalgo}, \citenamefont {Hausmann},\ and\
  \citenamefont {Hidalgo}}]{hausmann.country.2010}%
  \BibitemOpen
  \bibfield  {author} {\bibinfo {author} {\bibfnamefont {R.}~\bibnamefont
  {Hausmann}}, \bibinfo {author} {\bibfnamefont {C.~A.}\ \bibnamefont
  {Hidalgo}}, \bibinfo {author} {\bibfnamefont {R.}~\bibnamefont {Hausmann}}, \
  and\ \bibinfo {author} {\bibfnamefont {C.~A.}\ \bibnamefont {Hidalgo}},\
  } { (\bibinfo {year} {2010})}%
\bibitem [{\citenamefont {Zhang}\ and\ \citenamefont
  {Yu}(2010)}]{zhang.allometric.2010}%
  \BibitemOpen
  \bibfield  {author} {\bibinfo {author} {\bibfnamefont {J.}~\bibnamefont
  {Zhang}}\ and\ \bibinfo {author} {\bibfnamefont {T.}~\bibnamefont {Yu}},\ }
   {\bibfield  {journal} {\bibinfo
  {journal} {Physica A: Statistical Mechanics and its Applications},\ }\textbf
  {\bibinfo {volume} {389}},\ \bibinfo {pages} {4887} (\bibinfo {year}
  {2010})},\ ISSN \bibinfo {issn} {0378-4371}%
\bibitem [{\citenamefont {Sözen}\ and\ \citenamefont
  {Arcaklioglu}(2007)}]{sozen.prediction.2007}%
  \BibitemOpen
  \bibfield  {author} {\bibinfo {author} {\bibfnamefont {A.}~\bibnamefont
  {Sözen}}\ and\ \bibinfo {author} {\bibfnamefont {E.}~\bibnamefont
  {Arcaklioglu}},\ } {\bibfield  {journal}
  {\bibinfo  {journal} {Energy Policy},\ }\textbf {\bibinfo {volume} {35}},\
  \bibinfo {pages} {4981} (\bibinfo {year} {2007})},\ ISSN \bibinfo {issn}
  {0301-4215}%
\bibitem [{\citenamefont {Mozumder}\ and\ \citenamefont
  {Marathe}(2007)}]{mozumder.causality.2007}%
  \BibitemOpen
  \bibfield  {author} {\bibinfo {author} {\bibfnamefont {P.}~\bibnamefont
  {Mozumder}}\ and\ \bibinfo {author} {\bibfnamefont {A.}~\bibnamefont
  {Marathe}},\ } {\bibfield  {journal}
  {\bibinfo  {journal} {Energy Policy},\ }\textbf {\bibinfo {volume} {35}},\
  \bibinfo {pages} {395} (\bibinfo {year} {2007})},\ ISSN \bibinfo {issn}
  {0301-4215}%
\bibitem [{\citenamefont {Camba-Mendez}\ \emph {et~al.}(2001)\citenamefont
  {Camba-Mendez}, \citenamefont {Kapetanios}, \citenamefont {Smith},\ and\
  \citenamefont {Weale}}]{camba-mendez.automatic.2001}%
  \BibitemOpen
  \bibfield  {author} {\bibinfo {author} {\bibfnamefont {G.}~\bibnamefont
  {Camba-Mendez}}, \bibinfo {author} {\bibfnamefont {G.}~\bibnamefont
  {Kapetanios}}, \bibinfo {author} {\bibfnamefont {R.~J.}\ \bibnamefont
  {Smith}}, \ and\ \bibinfo {author} {\bibfnamefont {M.~R.}\ \bibnamefont
  {Weale}},\ } {\bibfield  {journal} {\bibinfo
  {journal} {Econometrics Journal},\ }\textbf {\bibinfo {volume} {4}},\
  \bibinfo {pages} {S56–S90} (\bibinfo {year} {2001})},\ ISSN \bibinfo {issn}
  {1368-{423X}}%
\bibitem [{\citenamefont {Haanstra}\ \emph {et~al.}(1985)\citenamefont
  {Haanstra}, \citenamefont {Doelman},\ and\ \citenamefont
  {Voshaar}}]{haanstra.use.1985}%
  \BibitemOpen
  \bibfield  {author} {\bibinfo {author} {\bibfnamefont {L.}~\bibnamefont
  {Haanstra}}, \bibinfo {author} {\bibfnamefont {P.}~\bibnamefont {Doelman}}, \
  and\ \bibinfo {author} {\bibfnamefont {J.~H.~O.}\ \bibnamefont {Voshaar}},\
  } {\bibfield  {journal} {\bibinfo  {journal} {Plant
  and Soil},\ }\textbf {\bibinfo {volume} {84}},\ \bibinfo {pages} {293}
  (\bibinfo {year} {1985})},\ ISSN \bibinfo {issn} {0032-{079X},
  1573-5036}%
\bibitem [{\citenamefont {Zwietering}\ \emph {et~al.}(1990)\citenamefont
  {Zwietering}, \citenamefont {Jongenburger}, \citenamefont {Rombouts},\ and\
  \citenamefont {van~'t Riet}}]{zwietering.modeling.1990}%
  \BibitemOpen
  \bibfield  {author} {\bibinfo {author} {\bibfnamefont {M.~H.}\ \bibnamefont
  {Zwietering}}, \bibinfo {author} {\bibfnamefont {I.}~\bibnamefont
  {Jongenburger}}, \bibinfo {author} {\bibfnamefont {F.~M.}\ \bibnamefont
  {Rombouts}}, \ and\ \bibinfo {author} {\bibfnamefont {K.}~\bibnamefont
  {van~'t Riet}},\ }{\bibfield  {journal} {\bibinfo  {journal} {Applied and Environmental
  Microbiology},\ }\textbf {\bibinfo {volume} {56}},\ \bibinfo {pages} {1875}
  (\bibinfo {year} {1990})},\ ISSN \bibinfo {issn} {0099-2240},\ \bibinfo
  {note} {{PMID:} 16348228 {PMCID:} {PMC184525}}%
\bibitem [{\citenamefont {Northam}(1979)}]{northam.urban.1979}%
  \BibitemOpen
  \bibfield  {author} {\bibinfo {author} {\bibfnamefont {R.~M.}\ \bibnamefont
  {Northam}},\ }{{\selectlanguage {english}\emph {\bibinfo {title}
  {Urban geography}}}}\ (\bibinfo  {publisher} {Wiley},\ \bibinfo {year}
  {1979})\ ISBN \bibinfo {isbn} {9780471032922}%
\bibitem [{\citenamefont {Hubbell}(2008)}]{hubbell.unified.2008}%
  \BibitemOpen
  \bibfield  {author} {\bibinfo {author} {\bibfnamefont {S.~P.}\ \bibnamefont
  {Hubbell}},\ }{{\selectlanguage {english}\emph {\bibinfo {title}
  {The Unified Neutral Theory of Biodiversity and Biogeography ({MPB-32)}}}}}\
  (\bibinfo  {publisher} {Princeton University Press},\ \bibinfo {year}
  {2008})\ ISBN \bibinfo {isbn} {9780691021287}%
\bibitem [{\citenamefont {Gaston}\ \emph {et~al.}(2004)\citenamefont {Gaston},
  \citenamefont {Spicer}, \citenamefont {Gaston},\ and\ \citenamefont
  {Spicer}}]{gaston.biodiversity.2004}%
  \BibitemOpen
  \bibfield  {author} {\bibinfo {author} {\bibfnamefont {K.~J.}\ \bibnamefont
  {Gaston}}, \bibinfo {author} {\bibfnamefont {J.~I.}\ \bibnamefont {Spicer}},
  \bibinfo {author} {\bibfnamefont {K.~J.}\ \bibnamefont {Gaston}}, \ and\
  \bibinfo {author} {\bibfnamefont {J.~I.}\ \bibnamefont {Spicer}},\ }{ (\bibinfo {year}
  {2004})}%
\bibitem [{\citenamefont {Mulder}\ \emph {et~al.}(2001)\citenamefont {Mulder},
  \citenamefont {Uliassi},\ and\ \citenamefont {Doak}}]{mulder.physical.2001}%
  \BibitemOpen
  \bibfield  {author} {\bibinfo {author} {\bibfnamefont {C.~P.~H.}\
  \bibnamefont {Mulder}}, \bibinfo {author} {\bibfnamefont {D.~D.}\
  \bibnamefont {Uliassi}}, \ and\ \bibinfo {author} {\bibfnamefont {D.~F.}\
  \bibnamefont {Doak}},\ } {\bibfield  {journal}
  {\bibinfo  {journal} {Proceedings of the National Academy of Sciences},\
  }\textbf {\bibinfo {volume} {98}},\ \bibinfo {pages} {6704} (\bibinfo {year}
  {2001})},\ ISSN \bibinfo {issn} {0027-8424, 1091-6490}%
\bibitem [{\citenamefont {Loreau}\ \emph {et~al.}(2001)\citenamefont {Loreau},
  \citenamefont {Naeem}, \citenamefont {Inchausti}, \citenamefont {Bengtsson},
  \citenamefont {Grime}, \citenamefont {Hector}, \citenamefont {Hooper},
  \citenamefont {Huston}, \citenamefont {Raffaelli}, \citenamefont {Schmid},
  \citenamefont {Tilman},\ and\ \citenamefont
  {Wardle}}]{loreau.biodiversity.2001}%
  \BibitemOpen
  \bibfield  {author} {\bibinfo {author} {\bibfnamefont {M.}~\bibnamefont
  {Loreau}}, \bibinfo {author} {\bibfnamefont {S.}~\bibnamefont {Naeem}},
  \bibinfo {author} {\bibfnamefont {P.}~\bibnamefont {Inchausti}}, \bibinfo
  {author} {\bibfnamefont {J.}~\bibnamefont {Bengtsson}}, \bibinfo {author}
  {\bibfnamefont {J.~P.}\ \bibnamefont {Grime}}, \bibinfo {author}
  {\bibfnamefont {A.}~\bibnamefont {Hector}}, \bibinfo {author} {\bibfnamefont
  {D.~U.}\ \bibnamefont {Hooper}}, \bibinfo {author} {\bibfnamefont {M.~A.}\
  \bibnamefont {Huston}}, \bibinfo {author} {\bibfnamefont {D.}~\bibnamefont
  {Raffaelli}}, \bibinfo {author} {\bibfnamefont {B.}~\bibnamefont {Schmid}},
  \bibinfo {author} {\bibfnamefont {D.}~\bibnamefont {Tilman}}, \ and\ \bibinfo
  {author} {\bibfnamefont {D.~A.}\ \bibnamefont {Wardle}},\ } {\bibfield  {journal} {\bibinfo  {journal}
  {Science},\ }\textbf {\bibinfo {volume} {294}},\ \bibinfo {pages} {804}
  (\bibinfo {year} {2001})},\ ISSN \bibinfo {issn} {0036-8075,
  1095-9203}%
\bibitem [{\citenamefont {He}\ and\ \citenamefont
  {Legendre}(1996)}]{he.species-area.1996}%
  \BibitemOpen
  \bibfield  {author} {\bibinfo {author} {\bibfnamefont {F.}~\bibnamefont
  {He}}\ and\ \bibinfo {author} {\bibfnamefont {P.}~\bibnamefont {Legendre}},\
  } {\bibfield  {journal} {\bibinfo  {journal} {The
  American Naturalist},\ }\textbf {\bibinfo {volume} {148}},\ \bibinfo {pages}
  {719} (\bibinfo {year} {1996})},\ ISSN \bibinfo {issn} {0003-0147,
  1537-5323}%
\bibitem [{\citenamefont {He}\ \emph {et~al.}(1996)\citenamefont {He},
  \citenamefont {Legendre},\ and\ \citenamefont
  {{LaFrankie}}}]{he.spatial.1996}%
  \BibitemOpen
  \bibfield  {author} {\bibinfo {author} {\bibfnamefont {F.}~\bibnamefont
  {He}}, \bibinfo {author} {\bibfnamefont {P.}~\bibnamefont {Legendre}}, \ and\
  \bibinfo {author} {\bibfnamefont {J.}~\bibnamefont {{LaFrankie}}},\ } {\bibfield  {journal} {\bibinfo  {journal}
  {Journal of Biogeography},\ }\textbf {\bibinfo {volume} {23}},\ \bibinfo
  {pages} {57-64} (\bibinfo {year} {1996})},\ ISSN \bibinfo {issn}
  {1365-2699}\%
\bibitem [{\citenamefont {Wei}\ \emph {et~al.}(2010)\citenamefont {Wei},
  \citenamefont {Li}, \citenamefont {Walther}, \citenamefont {Ye},
  \citenamefont {Huang}, \citenamefont {Cao}, \citenamefont {Lian},
  \citenamefont {Wang},\ and\ \citenamefont {Chen}}]{wei.comparative.2010}%
  \BibitemOpen
  \bibfield  {author} {\bibinfo {author} {\bibfnamefont {S.-g.}\ \bibnamefont
  {Wei}}, \bibinfo {author} {\bibfnamefont {L.}~\bibnamefont {Li}}, \bibinfo
  {author} {\bibfnamefont {B.~A.}\ \bibnamefont {Walther}}, \bibinfo {author}
  {\bibfnamefont {W.-h.}\ \bibnamefont {Ye}}, \bibinfo {author} {\bibfnamefont
  {Z.-l.}\ \bibnamefont {Huang}}, \bibinfo {author} {\bibfnamefont {H.-l.}\
  \bibnamefont {Cao}}, \bibinfo {author} {\bibfnamefont {J.-Y.}\ \bibnamefont
  {Lian}}, \bibinfo {author} {\bibfnamefont {Z.-G.}\ \bibnamefont {Wang}}, \
  and\ \bibinfo {author} {\bibfnamefont {Y.-Y.}\ \bibnamefont {Chen}},\ } {\bibfield  {journal} {\bibinfo  {journal}
  {Ecological Research},\ }\textbf {\bibinfo {volume} {25}},\ \bibinfo {pages}
  {93} (\bibinfo {year} {2010})},\ ISSN \bibinfo {issn} {0912-3814,
  1440-1703}%
\bibitem [{\citenamefont {Scheiner}(2003)}]{scheiner.six.2003}%
  \BibitemOpen
  \bibfield  {author} {\bibinfo {author} {\bibfnamefont {S.~M.}\ \bibnamefont
  {Scheiner}},\ } {\bibfield  {journal}
  {\bibinfo  {journal} {Global Ecology and Biogeography},\ }\textbf {\bibinfo
  {volume} {12}},\ \bibinfo {pages} {441-447} (\bibinfo {year} {2003})},\
  ISSN \bibinfo {issn} {1466-8238}%
\bibitem [{\citenamefont {Hidalgo}\ \emph {et~al.}(2007)\citenamefont
  {Hidalgo}, \citenamefont {Klinger}, \citenamefont {Barabási},\ and\
  \citenamefont {Hausmann}}]{hidalgo.product.2007}%
  \BibitemOpen
  \bibfield  {author} {\bibinfo {author} {\bibfnamefont {C.~A.}\ \bibnamefont
  {Hidalgo}}, \bibinfo {author} {\bibfnamefont {B.}~\bibnamefont {Klinger}},
  \bibinfo {author} {\bibfnamefont {A.-L.}\ \bibnamefont {Barabási}}, \ and\
  \bibinfo {author} {\bibfnamefont {R.}~\bibnamefont {Hausmann}},\ } {\bibfield  {journal} {\bibinfo  {journal}
  {Science},\ }\textbf {\bibinfo {volume} {317}},\ \bibinfo {pages} {482}
  (\bibinfo {year} {2007})},\ ISSN \bibinfo {issn} {0036-8075,
  1095-9203}%
\bibitem [{\citenamefont {Ausloos}\ and\ \citenamefont
  {Lambiotte}(2007)}]{ausloos.clusters.2007}%
  \BibitemOpen
  \bibfield  {author} {\bibinfo {author} {\bibfnamefont {M.}~\bibnamefont
  {Ausloos}}\ and\ \bibinfo {author} {\bibfnamefont {R.}~\bibnamefont
  {Lambiotte}},\ } {\bibfield  {journal}
  {\bibinfo  {journal} {Physica A: Statistical Mechanics and its
  Applications},\ }\textbf {\bibinfo {volume} {382}},\ \bibinfo {pages} {16}
  (\bibinfo {year} {2007})},\ ISSN \bibinfo {issn} {0378-4371}%
\end{thebibliography}%

\end{document}